%% file: Merge.tex
\documentclass[aip,apl,preprint]{revtex4-2}
\usepackage{xspace} 
\usepackage{makecell} 
\usepackage{siunitx, mhchem}
\usepackage{color,soul}
\usepackage{graphicx}
\usepackage{bm}        
\usepackage{amssymb}   
\usepackage{amsmath,mathtools} 
\usepackage{breqn} 
\usepackage{multirow}
\usepackage{array}
\usepackage{booktabs}
\usepackage[export]{adjustbox}
\usepackage{textcomp}
\usepackage[normalem]{ulem} 
\usepackage{gensymb}
\newcommand{\mos}{MoS$_2$\xspace}

\begin{document}
\include{Manuscript}
\include{Supplemetary}
\end{document}

%% file: Manuscript.tex

\title{Influence of Defects on the Valley Polarization Properties of Monolayer MoS$_{2}$ Grown by Chemical Vapor Deposition}

\author{Faiha\ Mujeeb}
\email{17faihamujeeb@gmail.com}
\author{Poulab\ Chakrabarti}
\author{Vikram\ Mahamiya}
\author{Alok\ Shukla}
\author{Subhabrata\ Dhar}
\email{dhar@phy.iitb.ac.in}
\affiliation{Department of Physics$,$ Indian Institute of Technology Bombay$,$ Powai$,$ Mumbai 400076$,$ India}

\begin{abstract}
\textbf{Abstract}: Here, the underlying mechanisms behind  valley de-polarization is investigated in chemical vapor deposited 1L-\mos. Temperature dependent polarization resolved photoluminescence spectroscopy was carried out on as-grown, transferred and capped samples. It has been found that the momentum scattering of the excitons due to the sulfur-vacancies attached with air-molecule defects have strong influence in valley de-polarization process. Our study reveals that at sufficiently low densities of such defects and temperatures, long range electron-hole exchange mediated intervalley transfer due to momentum scattering via Maialle-Silva-Sham (MSS) mechanism of excitons is indeed the most dominant spin-flip process as suggested by the theory\cite{WuMSS}. Rate of momentum scattering  of the excitons due to these defects is found to be proportional to the cube root of the density of the defects. Intervalley transfer process of excitons involving $\Gamma$-valley also has significance in valley de-polarization process specially when the layer has tensile strain or high density of $V_S$ defects as  these perturbations reduces $K$ to $\Gamma$-energy separation. Band-structural calculations carried out within density functional theory framework validate this finding. Experimental results further suggest that exchange interactions with the physisorbed air molecules can also result in the intervalley spin-flip scattering  of the excitons and this process give an important contribution to valley depolarization specially at the strong scattering regime. 	
\end{abstract}

	\maketitle 
		
	
Two-dimensional transition metal dichalcogenides (TMDs) offer valley degree of freedom, which can be exploited to design  next-generation valley based electronics or $\lq$valleytronics' \cite{xiao2012coupled}. The broken inversion symmetry, together with strong spin-orbit coupling, results in the valley-dependent optical selection rules in monolayer (1L)-\mos. This property  enables an exciton to sustain its valley character throughout the time of its existence. In fact, as high as 100$\%$ valley polarization has been reported in exfoliated 1L-\mos samples\cite{zhu2011giant,mak2012control,yao2008valley,xiao2012coupled,cao2012valley}. Whereas, 1L-\mos films grown by chemical vapour deposition (CVD) technique, which is frequently used to grow large area films on different substrates, show only moderate values of polarization (less than 50$\%$)\cite{shree2019high}. Since large area coverage of the monolayer film has to be ensured for any practical application of the material, it is imperative to understand the reason for moderation of valley polarization in CVD grown 1L-\mos. Note that the optical and electrical properties of CVD grown layers often suffer from the presence of a high density of sulfur vacancy defects ($V_{S}$) and the residual strain\cite{tongay2013broad,chow2015defect,nan2014strong,tailoring,miller_physisorbedmolecule,amani2014growth,zhu2013strain,chow2015defect}. Since the valley and spin properties are closely related to the crystal symmetry, both the strain\cite{zhu2013strain,chakrabarti2022enhancement} and the defects\cite{kim_ChemiDoping_VP,klein_defect_VP,mupparapu_Defect_VP,Nihit_LiDoping} are expected to have certain impacts on the valley polarization (VP) property of 1L-\mos grown by CVD technique. It has indeed been experimentally  demonstrated that VP decreases with increasing tensile strain in the 1L-\mos\cite{zhu2013strain,chakrabarti2022enhancement}. This has been explained in terms of longitudinal acoustic (LA) phonon assisted intervalley scattering of the excitons via $\Gamma$ valley as  the $K$ to $\Gamma$-valley energy separation decreases with the increase of biaxial tensile strain\cite{shi2013_StrainDFT,scalise2014_strainDFT}. However, the underlying mechanism through which defects govern VP in this system is yet to be systematically investigated.

In an ideal scenario, bright excitons generated in one of the $K$-valleys through  circularly polarized ($\sigma^+$ or $\sigma^-$ polarization) photons are expected to stay in the same valley until recombination. One may think that intervalley phonon scattering along with spin flipping of both electron and hole are necessary to transfer a bright exciton between $K$ to $K'$-valleys. However,  such processes are rare because neither D’yakonov-Perel’ (DP) nor Elliott-Yafet (EY) mechanism can result in spin relaxation of electrons/holes as the out-of-plane spin component is conserved for both the carriers\cite{xiao2012coupled,wang2014intrinsic,kormanyos2013monolayer,ochoa2013spin}. But in reality,  excitons do move between $K$ to $K'$-valleys and in certain cases, the de-polarization rate is shown to be extremely fast even in exfoliated 1L-\mos samples\cite{lagarde2014carrier,mai2014many,wang2013valley}. A recent theory suggests that long range part of the electron-hole exchange interaction can virtually transfer excitons between $K$ to $K'$-valleys\cite{WuMSS} without directly involving any phonon. In this process, excitons can experience in-plane effective magnetic field $\Omega(P_{ex})$ that depends upon its in-plane centre of mass momentum  $P_{ex}$. Precession of the exciton total angular momentum about $\Omega(P_{ex})$ can cause valley de-polarization due to inhomogeneous broadening. Exciton momentum scattering rate  can influence its spin scattering rate through Maialle-Silva-Sham (MSS) mechanism, which has similar characteristics as the DY process for the electrons and holes.  In the weak scattering regime, the spin scattering rate is proportional to the momentum scattering rate, while the two rates follow the inverse relationship in the strong scattering regime. Presence of defects can thus influence the momentum relaxation rate of the excitons and hence can affect the valley de-polarization. 

Here we explore the influence of sulfur vacancy related defects on the valley polarization property of CVD grown 1L-MoS$_2$. Our study reveals that momentum scattering of the excitons due to the sulfur-vacancies, which are physisorbed with air-molecules, influence the intervalley spin-flip transition rate of the excitons and hence the valley de-polarization process.  Both weak and strong scattering regimes of the intervalley excitonic transfer processes could indeed be identified from the  dependence of the degree of valley polarization on the defect concentration and the temperature, validating the MSS mechanism. It has been found that in the presence of biaxial tensile strain, high defect densities and/or at sufficiently high temperatures, (LA) phonon assisted intervalley scattering  via $\Gamma$ valley becomes important. This has been corroborated by ab-initio band-structural calculations.

Three types of samples were used for the study; CVD-grown 1L-MoS$_{2}$ films on sapphire substrates [sample M1],  1L-MoS$_{2}$ films-on-sapphire capped by the deposits from hBN pellets using pulsed laser deposition(PLD) [sample M2], and CVD grown 1L-MoS$_{2}$ films transferred onto SiO$_{2}$/Si substrates using a polystyrene(PS) based surface energy-assisted transfer procedure [sample M3].  More details about the growth, transfer process and characterizations  of these samples can be found in the supplementary. Photoluminescence (PL) and polarization-resolved PL studies were conducted keeping the samples in a liquid nitrogen cryostat.  Measurements were carried out  in backscattering configuration within a microscope set-up equipped with a 50X long working distance objective (NA= 0.5). For PL, a 532~nm diode laser was used as excitation source.  For polarization-resolved PL, an achromatic quarter wave-plate was used to produce circularly polarized lights ($\sigma^{-/+}$) from the linearly  polarized HeNe (632.8 nm) laser. A combination of a separate achromatic quarter wave-plate and a Glan-Tylor analyzer was placed before spectrometer entrance slit to select between $\sigma^{-}$ and $\sigma^{+}$ emitted photons. The spectra were recorded using a 0.55~m focal length monochromator equipped with Peltier cooled CCD detector. To avoid Joule heating, excitation intensity was kept at 150 $\mu$w on a spot diameter of $\sim$5 $\mu$m. 

\begin{figure}[htb]
	\includegraphics[scale=1.2]{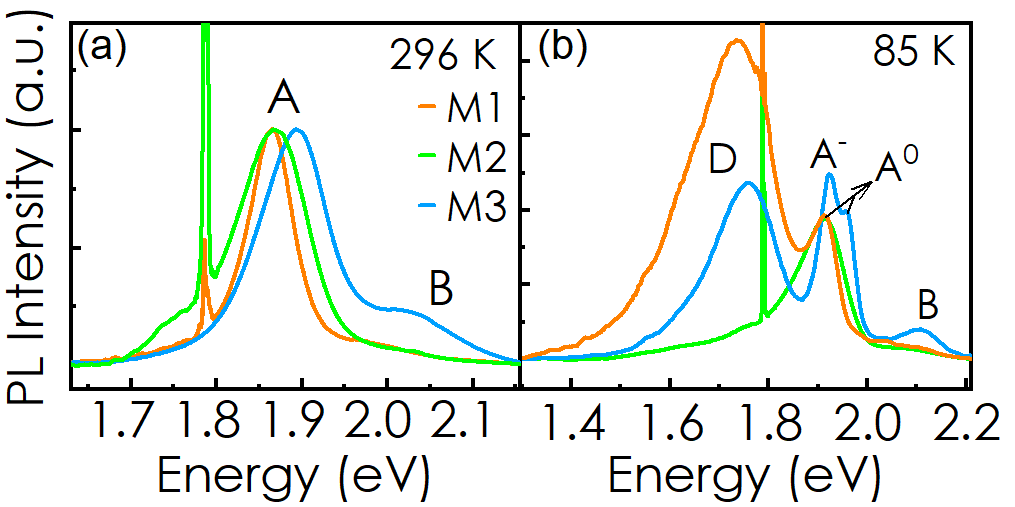}
	\caption{Normalized (with respect to the A excitonic feature) photoluminescence spectra recorded with 532~nm laser excitation for different samples at (a) room temperature and (b) 85 K.}
	\label{fig:fig1plrt}
	\vspace*{-\baselineskip}
\end{figure}
In the room temperature PL spectra recorded with 532~nm laser excitation for the three samples as shown in Fig.~\ref{fig:fig1plrt}(a), A-excitonic features are found at almost the same energy positions for the as-grown and the capped samples, whereas the feature appears at a higher energy position for the transferred sample. This blue shift implies the release of the tensile strain after transfer of the monolayer from the sapphire to the amorphous SiO$_2$/Si substrate \cite{amani2014growth,zhu2013strain}. Note that the as-grown 1L-\mos layer on sapphire is expected to be under a tensile biaxial strain due to the mismatch in the thermal expansion coefficient and/or lattice constant\cite{ayers2016heteroepitaxy,dumcenco2015large,ye2014germanium}. 
The broad luminescence feature (D) appearing at $\sim$ 1.75 eV in Fig.~\ref{fig:fig1plrt}(b), where 85~K PL spectra are compared, can be attributed  to those $V_S$ sites where  air molecules, such as oxygen and water, are physisorbed\cite{tailoring,chakrabarti2022enhancement,tongay2013defects}. Evidently, the intensity of D peak is significantly less in the transferred sample M3 as compared to that of the as-grown sample M1. D-peak is almost fully suppressed in the capped sample M2. Annealing followed by capping in the preparation process of samples M2 and M3 are found to be the reason for the reduction of D feature\cite{chakrabarti2022enhancement}. It is interesting to note that the trion peak is stronger than the excitonic feature in sample M3, implying a large enhancement of electron concentration, which can be attributed to polystyrene. As an aromatic hydrocarbon, PS has the potential to act as donors\cite{chakrabarti2022enhancement}. 

\begin{figure}[h!]
	\includegraphics[scale=1.2]{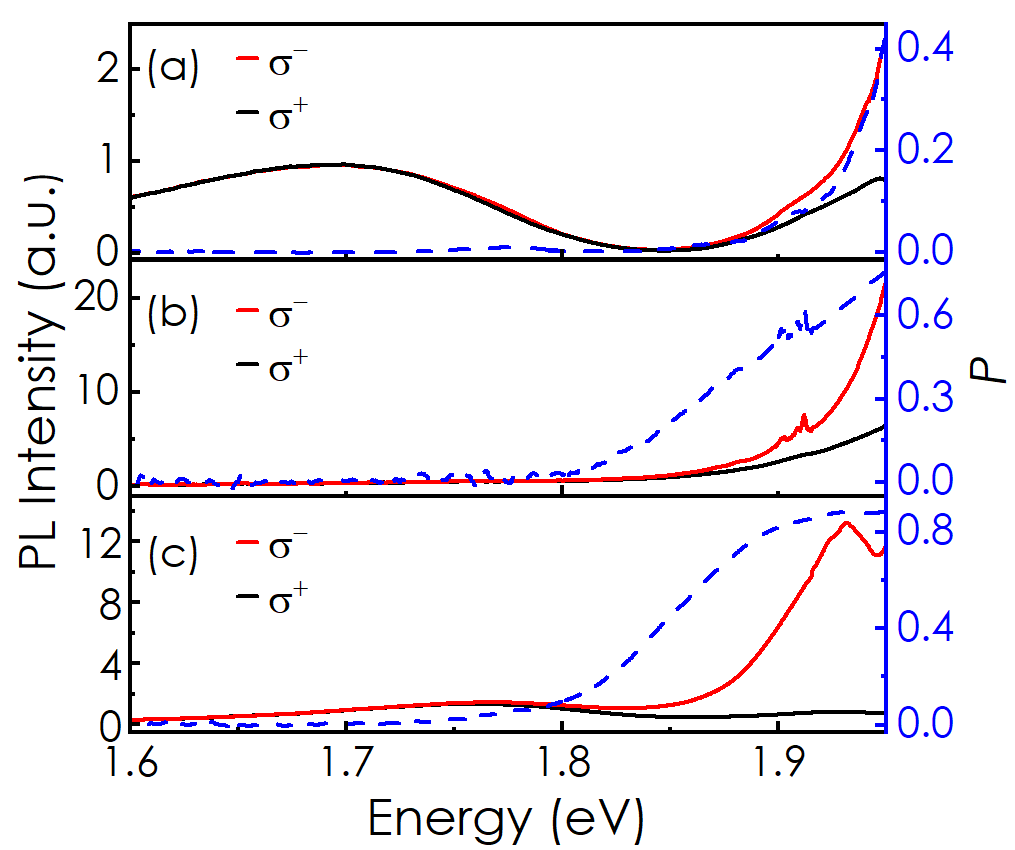}
	\caption{Circular polarization resolved PL spectra recorded at 85~K for the sample (a)M1, (b)M2 and (c)M3. Degree of polarization ($P$) is also plotted as a function of photon energy in these figures (dashed blue lines).}
	\label{fig:fig2}
	\vspace*{-\baselineskip}
\end{figure}

Fig.~\ref{fig:fig2} shows the polarization resolved PL spectra with $\sigma^{-}$ excitation recorded at 85~K on the three samples. Degree of valley polarization, which is defined as $P$ = $(I^{-}-I^{+})/(I^{-} + I^{+})$ with $I^{-/+}$ as the intensity of $\sigma^{-/+}$ light is also plotted as functions of photon energy in respective panels. Evidently, in all cases, polarization could only be observed at the A-exciton/trion features, while the D-band does not show any polarization at all. Note that D-feature is almost completely suppressed in the sample M2 and the sample shows higher $P$ than that is typically obtained in as-grown samples. This finding highlights the role of $V_S$-Air defects  in determining the valley polarization property of the material.  Interestingly, $P$ goes as high as 82$\%$ in case sample M3, where the intensity of the D-peak is significantly reduced as compared to that is generally found in as grown samples, but the reduction is not as much as it found  in sample M2. Observation of higher $P$ in sample M3 than M2 even when  $V_S$-Air defect density is larger in the former, can be attributed to the relaxation of biaxial strain in the \mos film after the transfer\cite{chakrabarti2022enhancement}. 

\begin{figure}
	\includegraphics[scale=1.2]{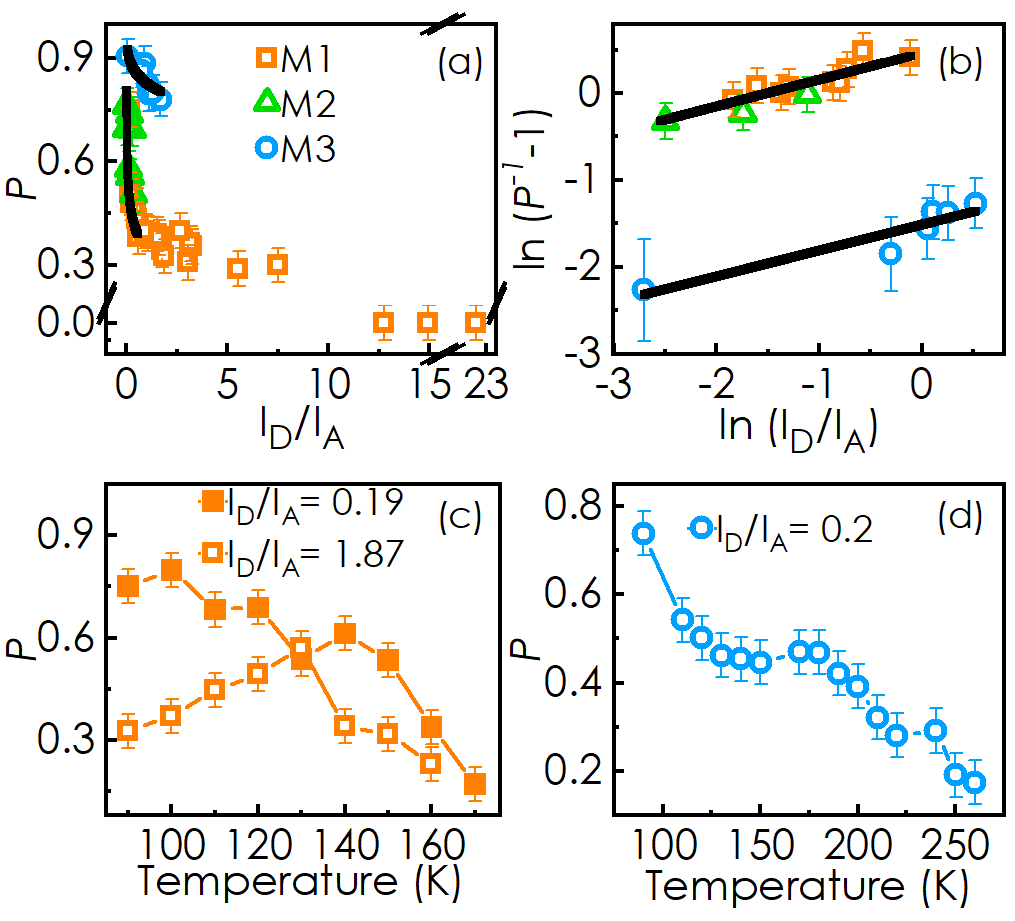}
	\caption{(a) Degree of circular polarization ($P$) as a function of $I_{D}/I_{A}$  obtained at various sampling points on different samples. (b) Plot of $\ln{(P^{-1}-1)}$ vs $\ln{(I_D/I_A)}$ for all the sampling points. Temperature dependence of $P$ for (c) as-grown samples with two different $I_D/I_A$ ratios and (d) the transferred sample.}
	\label{fig:fig3}
	\vspace*{-\baselineskip}
\end{figure}

Polarization resolve PL spectra are recorded at several spots on each sample. The relative intensity of the D-feature with respect to A-exciton complex $I_{D}/I_{A}$ at each sampling point can serve as a measure for the density of $V_S$-Air defects at that location. Note that in case of the as-grown samples, even though the ratio is found to vary significantly from  spot to spot, the position of both D- and A-features does not change much. In the case of capped and transferred samples, neither the $I_{D}/I_{A}$ ratio nor the peak positions show much spatial variation. Degree of polarization $P$ obtained at 85~K from various parts of these samples is plotted versus $I_{D}/I_{A}$ in Fig.~\ref{fig:fig3}(a). In case of the as-grown and the capped samples (M1 and M2), $P$ obtained at a fixed energy of 1.945~eV, while for the transferred sample, $P$ measured at the peak of the A-exciton/trion complex is used for the plot. Since the higher energy side of the PL feature corresponding to the A-exciton/trion complex can not be visible with the 633 nm (1.96~eV) excitation, $I_{D}/I_{A}$ ratio is obtained from the PL spectra recorded with 532 nm laser excitation at the same spot in all cases. Evidently, for the as-grown and the capped samples, all the data obey a trend of rapid initial decrease followed by a plateauing as $I_{D}/I_{A}$ ratio increases. Interestingly, beyond a certain $I_{D}/I_{A}$ ratio, $P$ suddenly drops to zero. Note that  the data obtained from the transferred sample stay clearly isolated from other data points in the plot. But, they also show a reduction as $I_{D}/I_{A}$ increases. These observations clearly demonstrate the correlation between the $V_S$-Air defects and $P$. 

Fig. \ref{fig:fig3}(c) compares the temperature ($T$) variation of $P$ recorded for an as-grown sample at two spots with different $I_{D}/I_{A}$ ratios. Interestingly, $P$ shows a monotonous decrease with the increase in temperature when  $I_{D}/I_{A}$ is only 0.19. While $P$ initially increases and then decreases with rising temperature for $I_{D}/I_{A}$ =1.87. We have investigated several spots with different $I_{D}/I_{A}$ ratios, and $P$ is found to consistently exhibit an initial trend of either reduction or enhancement with increasing $T$ depending upon whether $I_{D}/I_{A}$ is sufficiently low or high, respectively. Fig.~\ref{fig:fig3}(d) plots $P$ as a function of $T$ for the transferred sample M3 at a spot with  $I_{D}/I_{A}$ =0.2.  $P$, in this case, shows a reduction followed by a plateauing tendency as $T$ increases. Beyond $\sim$250~K, the polarization suddenly drops to zero.

Upon illumination with a circularly polarised light falling perpendicularly to the layer plane, A-excitons/trions are generated in one of the $K$-valleys depending upon the helicity of the incident light. Generated excitons can either be transferred to other non-equivalent $K$ valleys through inter-valley transition processes or can be captured by the $V_S$-Air defect centers before recombination. One can consider that the excitons are generated at a rate of $G$ in only one of the valleys (say $K$-valley). At the steady state condition, the population of excitons in the $K$, $K'$-valleys ($X$ and $X'$) and the defect sites ($X_D$) can be obtained in terms of $G$, total recombination (radiative plus non-radiative) rate $\gamma$ of the excitons, inter-valley relaxation rate $\gamma_{s}$ of the excitons, total recombination rate of the bound excitons $\gamma_{D}$, coefficient of transition of A-excitons to the defect bound state $\beta$ and the defect concentration $N_{D}$ by solving the rate equations. Considering $X_{D}/N_{D}$ $<<$ 1, polarization can be obtained as $P$ = $1/[1+ 2\gamma_s/(\gamma+\beta N_D)]$. One may further contemplate that the rate of recombination of A-excitons $\gamma$ is much higher than the rate of their capture at the defect sites $\beta N_{D}$. Polarization can then be expressed as $P$ = $1/[1+ 2\gamma_s/\gamma]$. More details of these calculations could be found in the supplementary. 

According to the theory proposed by Yu and Wu, inter-valley spin scattering rate $\gamma_s$ of the A-excitons should depend on the momentum scattering rate ($r_p$) of the excitons through Maialle-Silva-Sham (MSS) mechanism\cite{WuMSS}. It is quite reasonable to believe that the presence of air-molecules at the S-vacancy sites introduces certain additional local vibrational modes, which can take part in the momentum scattering of excitons. One may thus consider that $r_p \propto N^{\alpha}_D$, where $\alpha$ is a constant. In the weak scattering regime,  $\gamma_s \propto r_p$ and hence $\gamma_s = Q_s N^{\alpha}_D$, where $Q_s$ is a constant. One can also express defect concentration $N_D$ in terms of the intensity ratio $I_D/I_A$ as $N_D$ = $(\gamma^r \gamma_D/\beta \gamma^r_D)(I_D/I_A)$, where $\gamma^r$ and $\gamma^r_D$ are the radiative recombination rate of A-excitons/trions and the defect bound excitons, respectively (see supplementary). $P$ can now be given by $P = 1/[1+S (I_D/I_A)^{\alpha}]$, where $S$ = $2 (Q_s/\gamma)(\gamma^r \gamma_D/\beta\gamma^r_D)^{\alpha}$. Note that $\gamma^r$ and $\gamma^r_{D}$ are independent of the defect concentration. At low defect densities and sufficiently low temperatures, recombination of excitons takes place mostly through radiative pathways and hence $\gamma \approx \gamma^r$ and  $\gamma_D \approx \gamma^r_D$. $S$ can thus be treated as independent of $N_D$.  The rapid initial fall in $P$ versus $I_D/I_A$ plot shown in Fig. \ref{fig:fig3}(a) can now be explained. In Fig. \ref{fig:fig3}(b), $\ln{(P^{-1}-1)}$ is plotted versus $\ln{(I_D/I_A)}$ for all data points shown in Fig. \ref{fig:fig3}(a). Data obtained for sample M1 and M2  clearly follow a straight line. While those from sample M3 can be fitted with a separate but parallel straight line with a slope of $\alpha$ =0.33. This further establishes the validity of the $P$ versus $I_D/I_A$ relationship in explaining the experimental results in the weak scattering regime. It should be noted that MSS mechanism predicts $\gamma_s$ $\propto$ $r^{-1}_p$ in the strong scattering regime, meaning $P$ should increase with $r_p$, while $r_p$ is expected to increase with $T$. In \ref{fig:fig3}(c) and (d), the observation of the initial decrease and increase of $P$ with the rising temperature when $I_{D}/I_{A}$ is low and high, respectively, can be assigned to the weak and strong scattering regimes, respectively. 

At sufficiently high temperatures, $P$ has been found to decrease with increasing $T$ in all cases. This can be attributed to the increase of center of mass momentum of the excitons $p_{ex}$. Since the band gap of the material decreases with the increase of $T$, for the same photon energy of excitation, the probability of generation of excitons with higher $p_{ex}$ increases, and according to the theory\cite{WuMSS}, $\gamma_s$ increases with $p_{ex}$. Polarization data presented in Fig. \ref{fig:fig3}(a) show a plateauing tendency beyond $I_D/I_A$ $\approx$ 1.5 before abruptly dropping down to zero at $I_D/I_A$ $\approx$ 6. However, theory predicts $P$ to enhance with $I_D/I_A$ beyond a certain point as the defect density moves from the weak to strong scattering regime. This may indicate the presence of other competing mechanisms, which increase the excitonic spin relaxation rate with $I_D/I_A$. One of the possible candidates might be the exchange interaction of the excitons with the air molecules attached at the $V_S$-sites. Note that certain air-molecules, such as O$_2$, H$_2$O, possess magnetic moment\cite{pitzer_AdsobedO2}. Physisorption of such molecules at the  $V_S$-sites, can interact with the excitons through exchange coupling. Spin relaxation rate of the excitons due to these scattering processes is expected to be proportional to the density of these defects\cite{wu2010spin,koschel1975zone}. We believe that the sudden drop of $P$ to zero when plotted as a function of $I_D/I_A$ (at $\sim$ 6) in Fig.~\ref{fig:fig3}(a) is due to the change in the band structure as a result of the inclusion of a large density of disorder in the lattice at such a high defect concentration. In fact, the reduction of the valley polarization with the increase of disorder in 1L-\mos has been reported\cite{wangTightBinding}. 

\begin{figure}
	\includegraphics[scale=1.5]{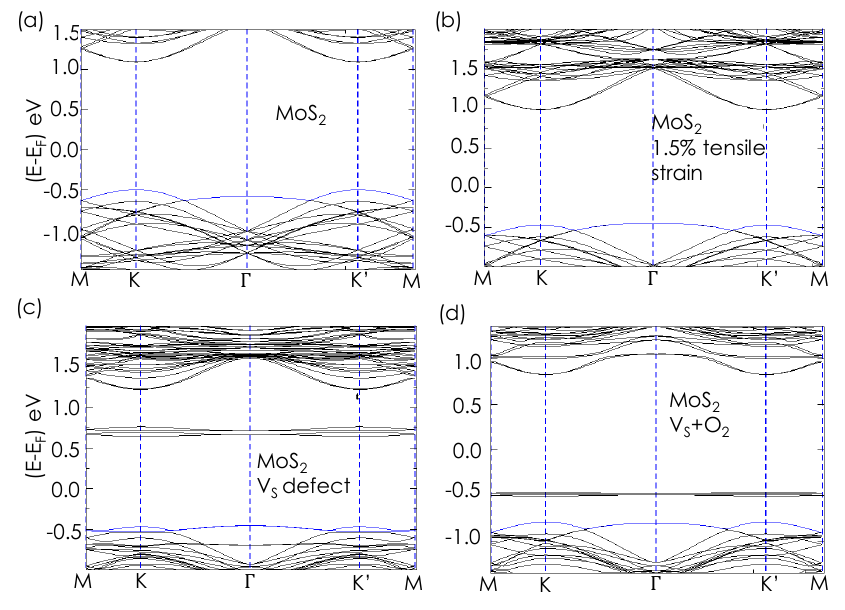}  
	\caption{Calculated band structures of 1L-\mos: when the film is (a) pristine, (b) biaxially strained (tensile), (c)unstrained but with bare $V_S$ defects and (d) unstrained but with $V_S$-O$_2$ defects.}
	\label{fig:dft}
	\vspace*{-\baselineskip}
\end{figure}

Here, we have carried out band structural calculations under the framework of density functional theory (DFT) to understand the effect of different perturbations such as $V_S$-formation, physi-adsorption of air-molecules with the $V_S$-sites and biaxial strain on the band structure of 1L-\mos. More details about the calculation can be found in the supplementary.

Calculated band structures for 1L-\mos, when the layer is pristine, biaxially strained (tensile), unstrained but with bare $V_S$ defects and unstrained but with $V_S$-O$_2$ defects, are compared in Fig.~\ref{fig:dft}. It is noticeable that as compared to the pristine layer, the energy separation between the $K/K'$- and the $\Gamma$-valley is reduced whenever the layer is either under a tensile biaxial strain or incorporated with the defects. Reduction of the energy gap can enhance the chance of the holes to transfer between the $K'$ and $K$-valleys via  $\Gamma$-valley through phonon assisted inter-valley processes. However, it has to be noted that the $z$-component of hole spin is still a good quantum number even for the $\Gamma$-valley. DY mechanism can not thus be a dominant process for spin relaxation in this path way\cite{WuMSS}. Rather, Elliott-Yafet (EY) mechanism should play more significant role in the hole spin relaxation process at the $\Gamma$-valley.

In conclusion, the momentum scattering of excitons due to $V_S$-air defects has been found to play a vital role in valley de-polarization process of CVD grown 1L-\mos. The study clearly demonstrates that at sufficiently low defect densities and temperatures, long range electron-hole exchange mediated transfer of excitons between $K$/$K'$-valleys indeed happens due to momentum scattering via MSS mechanism as theoretically proposed\cite{WuMSS}. Momentum scattering rate  of the excitons due to these defects comes out to be proportional to the cube root of the defect density. Intervalley transfer process of excitons involving $\Gamma$-valley also play substantial role specially when the layer has tensile strain or high density of $V_S$ defects as $K$ to $\Gamma$-energy separation decreases with these perturbations. The study further suggests that the exchange interaction between the excitons and the physisorbed air molecules can also lead to intervalley spin-flip scattering. Such processes also give substantial contribution to valley depolarization specially at strong scattering regime. 

\noindent\textbf{Supplementary:}
See supplementary material for the details of the sample preparation and the calculation of degree of polarization as a function of defect concentration.

\noindent\textbf{Acknowledgment:}
We acknowledge various experimental facilities provided by Sophisticated Analytical Instrument Facility (SAIF) and Industrial Research and Consultancy Centre (IRCC) of IIT Bombay.

\bibliographystyle{unsrt}
\bibliography{ref}


%% file: Supplemetary.tex




\noindent
{\huge Supplementary Information}
\\

\noindent
\textbf{{\Large{Influence of Defects on the Valley Polarization Properties of Monolayer MoS$_{2}$ Grown by Chemical Vapor Deposition}}}
\\
{\large Faiha Mujeeb, Poulab Chakrabarti, Subhabrata Dhar}\\
Department of Physics, Indian Institute of Technology Bombay, Powai, Mumbai-400076, India


\date{\today}
\maketitle


	\section{Growth of 1L-MoS$_{2}$ using Chemical Vapor Deposition}
1L-MoS$_{2}$ films were grown on double-sided polished c-plane sapphire using microcavity based chemical vapor deposition technique. High purity MoO$_{3}$ (99.5 $\%$) and S (99.7 $\%$) were used as precursors and Argon was used as a carrier gas. The MoO$_{3}$ (6 g) was filled inside a ceramic boat and placed at the center of the 2-inch furnace. Another boat was filled with S powder (350 g) and placed at one end. Three substrates were used, in which two were placed parallel to each other on top of the MoO$_{3}$ filled boat, and the third one was kept on top of both in the middle, and it is supported by two sapphire strips. The gap formed between the substrate and sapphire strips acts as a natural reactor cavity. More details about the growth method can be found elsewhere\cite{pkm}. The AFM image of the film grown on sapphire substrate is shown in Fig.~\ref{fig:afm}. The average height (after sampling at different locations of the image) is found to be 0.7 nm, which is typically reported for a 1L-MoS$_{2}$ film.

\begin{figure}[htb]
	\centering
	\includegraphics[scale=1]{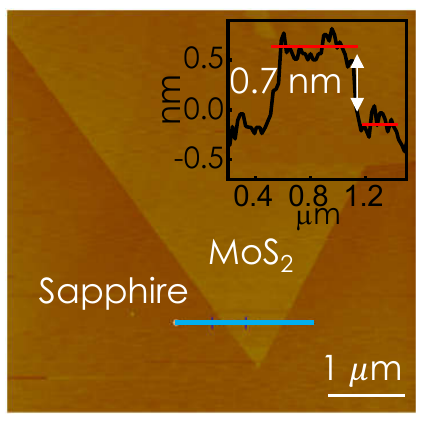}
	\caption{AFM image for the as-grown sample. The inset shows the height distribution profile along the blue line drawn across 1L-MoS$_{2}$- substrate boundary.}
	\label{fig:afm}
	\vspace*{-\baselineskip}
\end{figure}

\section{Transfer of 1L-MoS$_{2}$ onto SiO$_{2}$/Si Wafer}
1L-MoS$_{2}$ films are transferred from the sapphire substrates to  SiO$_{2}$/Si wafers using a surface energy assisted transfer procedure using Polystyrene (PS) as the carrier polymer\cite{gurarslan2014surface}. The PS solution was spin-coated on as-grown MoS$_{2}$/sapphire samples and then baked at 80- 90$^{\circ}$C for 35 min followed by 120$^{\circ}$C for 10 min.   MoS$_{2}$/PS assembly was gently poked at the edge, while water was dropwise added onto the film, which allows the water to go underneath the MoS$_{2}$. This helps MoS$_{2}$/PS assembly to be lifted from the sapphire substrate. The MoS$_{2}$/PS assembly was then transferred onto SiO$_{2}$/Si wafer. After transferring to the target substrate, it was again baked for 80- 90$^{\circ}$C for 35 minutes followed by 120$^{\circ}$C for 10 minutes. Finally, the PS layer is removed by rinsing it in toluene. The AFM image for the transferred sample is shown in Fig.~\ref{fig:afm}. The average height between the 1L-\mos and the substrate for the transferred sample is found to be of $\sim$3 nm with a rms roughness of 0.8 nm. This suggests that a $\sim$2.3 nm thick coating of PS is still remaining on top of the film even after rinsing in toluene for several times. 

\begin{figure}[htb]
	\centering
	\includegraphics[scale=1]{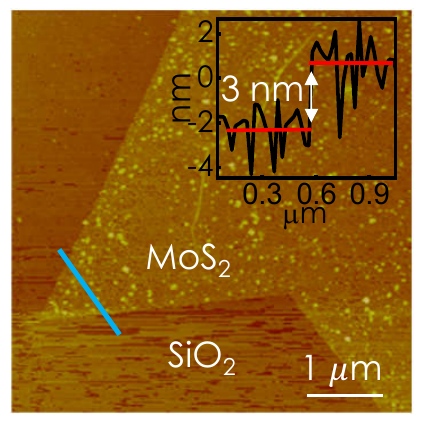}
	\caption{AFM image for the transferred sample. The inset shows the height distribution profile along the blue line drawn across 1L-MoS$_{2}$- substrate boundary.}
	\label{fig:afm}
	\vspace*{-\baselineskip}
\end{figure}

\section{Capping of 1L-MoS$_{2}$ using hBN Pellets}
The 1L-MoS$_{2}$ samples were vacuum annealed in a pulsed laser deposition chamber and capped using the deposit formed using the laser ablation of the h-BN pellet. A KrF excimer laser with a wavelength of 248 nm and pulse width of 25 ns was used to ablate the h-BN pellet. The energy density of the laser pulse was kept at 1.4 J cm$^{-2}$ at a frequency of 5 Hz. The base pressure of the chamber was measured to be less than 1 x 10$^{-5}$ mbar. The deposition is performed under N$_{2}$ atmosphere, keeping a pressure of 2 x 10$^{-2}$ mbar inside the chamber under a flow of 5N pure N$_{2}$ gas at 100 sccm. The substrate to target working distance was kept at 5 cm. The Scanning electron microscope (SEM) image of the deposited film is given in Fig.~\ref{fig:afm}(c). The thickness of the deposited film is found to be $\sim$ 22 nm using cross-sectional SEM, shown in the inset of Fig.~\ref{fig:afm}(c). 

\begin{figure}[htb]
	\centering
	\includegraphics[scale=1]{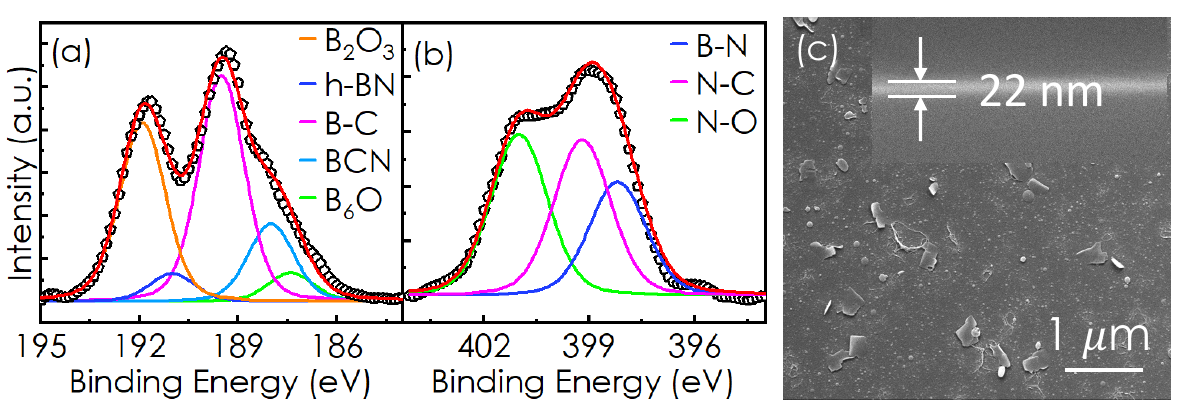}
	\caption{XPS core level spectra for (a) B 1s and (b) N 1s levels of the capped film. (c) SEM image of the deposited layer using BN pellets. The cross-section image is shown in the inset.}
	\label{fig:xps}
	\vspace*{-\baselineskip}
\end{figure}
The XPS spectra of both B 1s and N 1s levels of the deposited layer are given in Fig.\ref{fig:xps} (a,b), respectively. After a Shirley background subtraction, the spectra are deconvoluted using mixed Gaussian (80\%)-Lorentzian (20\%) functions. As shown in the Fig. \ref{fig:xps} (a), the B 1s spectrum is deconvoluted with five peaks attributed to B$_{2}$O$_{3}$ (192 eV)\cite{xps_B1s_192}, h-BN (191 eV)\cite{xps_B1s_191}, B-C (189.2 eV)\cite{xps_B1s_189}, BCN (188 eV)\cite{xps_B1s_188} and B$_{6}$O (187.4 eV)\cite{xps_B1s_187}. The N 1s spectrum is deconvoluted with three peaks corresponding to h-BN (398.2 eV)\cite{xps_N1S_398}, N-C (399.2 eV)\cite{xps_N1S_399} and N-O (401 eV)\cite{xps_N1S_401}.

\section{Raman spectra}
\begin{figure}[htb]
	\centering
	\includegraphics[scale=1]{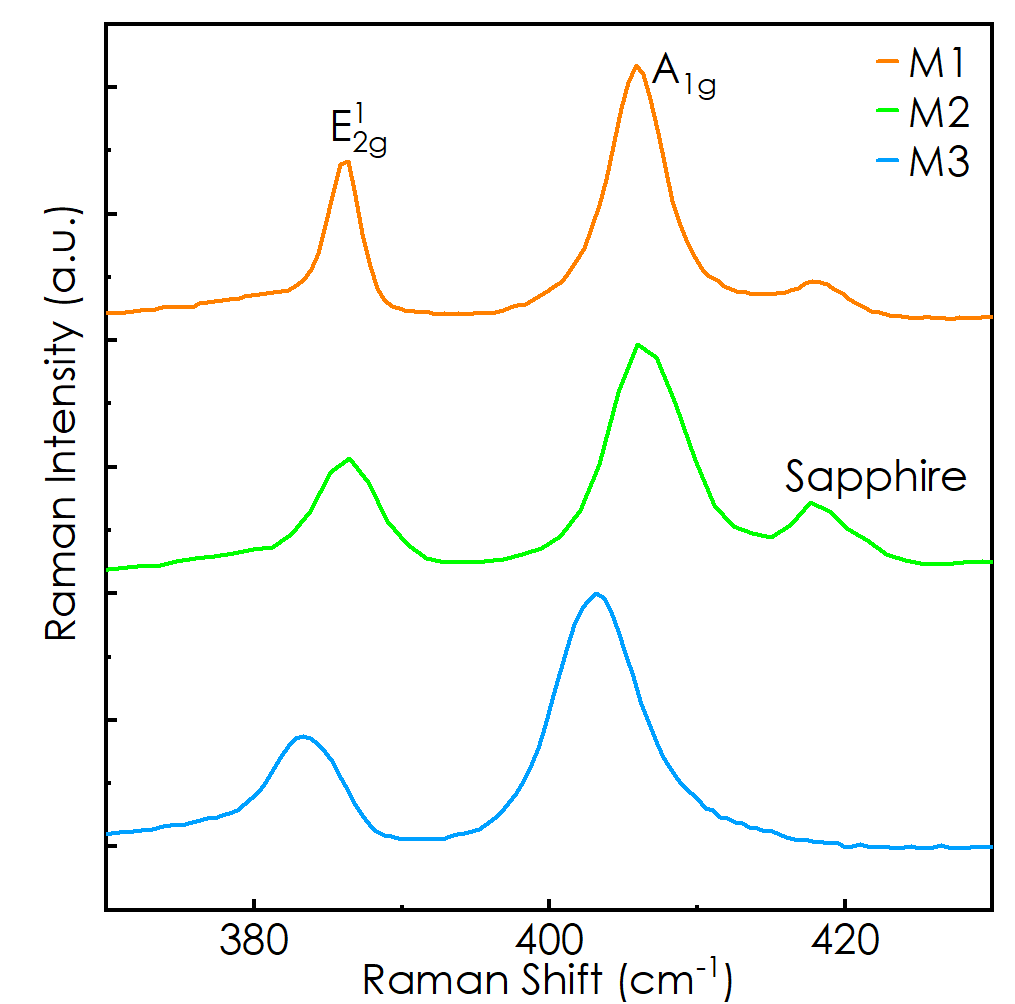}
	\caption{Room temperature Raman spectra normalized at A$_{1g}$ peak for as-grown (sample M1) and capped (sample M2) and transferred (sample M3) samples.}
	\label{fig:raman}
\end{figure}

Raman spectra on these samples were recorded at room temperature in back scattering geometry with 532 nm diode laser using Horiba JobinYvon HR800 confocal Raman spectrometer. Results are shown in Fig.~\ref{fig:raman}. In all cases, the characteristic in-plane E$^1_ {2g}$ and out-of-plane A$_{1g}$ vibrational modes for the zone-center phonons could be seen at $\sim$385 and $\sim$405 cm$^{-1}$, respectively. The separation  between the two featires, which serves as a good indicator of the layer thickness, comes out to be $\sim$ 20~cm$^{-1}$ for every sample. This further demonstrates monolayer nature of these \mos flakes.

\section{Dynamics of Optical Pumping of valley polarization in 1L-MoS$_{2}$}
The population of exciton at $K/K^{\prime}$ valleys ($X$/$X^{\prime}$) and bound excitons ($X_D$) can be written as, 
\begin{align}	
	\frac{dX}{dt}&=G-\gamma X-\beta(N_{D}-n_{D})X-(X-X^{\prime})\gamma_{s}\\
	\frac{dX^{\prime}}{dt}&=-\gamma X^{\prime}-\beta(N_{D}-n_{D})X^{\prime}-(X^{\prime}-X)\gamma_{s} \\
	\frac{dX_D}{dt}&=\beta(N_{D}-X_D)(X+X^{\prime})-\gamma_{D}X_D
\end{align}
Where $G$ is the pumping rate of neutral exciton at $K$ valley by left circularly polarized light. And, $\gamma$, $\gamma_{s}$ and $\gamma_{D}$ are total recombination rate of exciton, inter-valley relaxation rate and the recombination rate of bound exciton, respectively. $N_{D}$, $n_{D}$ and $\beta$ is the defect concentration in the sample, the population of neutral exciton bound to the defect state and the coefficient of transition of exciton from free to bound (defect) state. 

Under steady state condition, $\frac{dX(X^{\prime})}{dt}=\frac{dX_D}{dt}=0$.
\begin{equation}
	\beta(N_{D}-n_{D})(X+X^{\prime})-\gamma_{D}X_{D}=0
\end{equation}
Under $N_{D}>>X_{D}$, 
\begin{equation}
	\frac{X_{D}}{(X+X^{\prime})}=\frac{\beta N_{D}}{\gamma_{D}}
\end{equation}
Also,
\begin{equation}
	-\gamma X^{\prime}-\beta(N_{D}-n_{D})X^{\prime}-(X^{\prime}-X)\gamma_{s}=0
	\label{6}
\end{equation}
The equation \ref{6} can be rearranged to,
\begin{equation}
	\frac{X^{\prime}}{X}=\frac{1}{1+\frac{\gamma+\beta N_{D}}{\gamma_{s}}}
\end{equation}
The polarization helicity (P) can be expressed as,
\begin{align}
	P&=\frac{I^{-}-I^{+}}{I^{-}+I^{+}}= \frac{X-X^{\prime}}{X+X^{\prime}} \\
	&=\frac{1}{1+2\frac{\gamma_{s}}{\gamma+\beta N_{D}}}
\end{align}

Assuming $\gamma$ and $\gamma_{s}$ depends on the defect concentration $N_{D}$ as follows,
\begin{align}
	\gamma_{s}&=Q_{s}N^{\alpha}_{D} \\
	\gamma&=\gamma_{0}+Q_{A}N^{\alpha_{1}}_{D}
\end{align}
Where $\gamma_{0}$ is the rate of recombination when the sample with no defect, $Q_{s}$ and $Q_{A}$ are constants.
\begin{equation}
	P=\frac{1}{1+2\frac{Q_{s}N_{D}^{\alpha}}{\gamma_{0}+Q_{A}N_{D}^{\alpha_{1}}+\beta N_{D}}}
	\label{rho3}
\end{equation}
When the defect concentration is low, $N_{D}\rightarrow 0$, $\gamma_{0}>>Q_{A}N_{D}^{\alpha_{1}}+\beta N_{D}$, $\gamma_{0}\sim \gamma$ equation \ref{rho3} becomes,
\begin{equation}
	P=\frac{1}{1+2\frac{Q_{s}N_{D}^{\alpha}}{\gamma}}
	\label{P3}
\end{equation}

The PL intensity of D feature $I_{D}$, normalized on total exciton intensity $I_{A}=I^{K}_{A}+I^{K^{\prime}}_{A}$,
\begin{equation}
	\frac{I_{D}}{I_{A}}=\frac{X_{D}\gamma_{D}^{r}}{(X+X^{\prime})\gamma^{r}}=\frac{\beta N_{D}\gamma_{D}^{r}}{\gamma_{D}\gamma^{r}}
	\label{Iratio}
\end{equation}
Using equation \ref{Iratio}, equation \ref{P3} becomes,
\begin{equation}
	P=\frac{1}{1+S\left(\frac{I_{D}}{I_{A}}\right)^{\alpha}}
\end{equation}
Where, $S=\frac{2Q_{s}}{\gamma}\left(\frac{\gamma^{r}\gamma_{D}}{\beta \gamma_{D}^{r}}\right)^{\alpha}$

\section{Computational Details}
\begin{figure}[htb]
	\centering
	\includegraphics[scale=0.36]{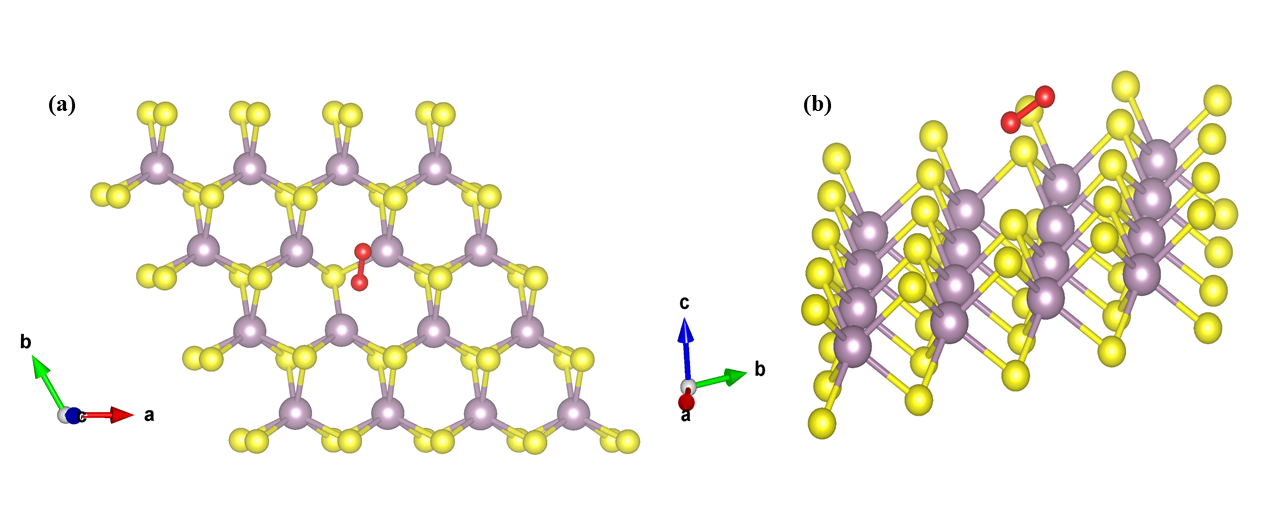}
	\caption{ Optimized structure of 4$\times$4 supercell of MoS$_2$ with one passivated sulfur vacancy (a) Top view and (b) Side view.}
	\label{fig:structure}
	\vspace*{-\baselineskip}
\end{figure}

We have carried out density functional theory (DFT) simulations\cite{DFT1} by employing the Vienna \textit{ab-initio} simulation package (VASP)\cite{DFT2,DFT3} to investigate the influence of strain and sulfur vacancies on the valley polarization properties of MoS$_2$. We have taken the Perdew-Burke-Ernzerhof (PBE)\cite{GGA,PAW} exchange-correlation functional along with the generalized gradient approximation and considered the plane wave basis expansion up to a kinetic energy cut-off of 450 eV for the calculations. The Monkhorst-pack \textit{k}-grids of 5$\times$5$\times$1 and 7$\times$7$\times$1 kpoints were taken for the geometry relaxation and density of states calculations, respectively. We apply a convergence limit of 0.02 eV/{\AA}  and 10$^{-5}$ eV for the calculations of Hellman-Feyman forces and total energy, respectively. The effects of spin-orbit coupling is incorporated and the van der Waals interactions are taken into account by employing Grimme's dispersion corrections of DFT-D3 type\cite{DFTD3}. A 4$\times$4 supercell of MoS$_2$ is considered for the simulations, and a vacuum space of 20 {\AA} is introduced along the z-direction to avoid the periodic interactions in our system. One sulfur vacancy is introduced in the 4$\times$4 supercell of MoS$_2$, and an oxygen molecule is adsorbed onto the sulfur vacancy. The top and side views of the relaxed structure of 4$\times4$ supercell of MoS$_2$ with one passivated sulfur vacancy are shown in Fig.~\ref{fig:structure}(a,b), respectively.

